# Shorter Distances between Papers over Time are Due to More Cross-Field References and Increased Citation Rate to Higher Impact Papers

Attila Varga

University of Arizona, School of Sociology, Social Sciences 400, P.O.Box 210027, Tucson, AZ 85721

**Keywords**: citation analysis, complex networks, science of science

**Abstract**: The exponential increase in the number of scientific publications raises the question of whether the sciences are expanding into a fractured structure, making cross-field communication difficult. On the other hand, scientists may be motivated to learn extensively across fields to enhance their innovative capacity, and this may offset the negative effects of fragmentation. Through an investigation of the distances within and clustering of cross-sectional citation networks, this study presents evidence that fields of science become more integrated over time. The average citation distance between papers published in the same year decreased from approximately 5.33 to 3.18 steps between 1950 and 2018. This observation is attributed to the growth of cross-field communication throughout the entire period as well as the growing importance of high impact papers to bridge networks in the same year. Three empirical findings support this conclusion. First, distances decreased between almost all disciplines throughout the time period. Second, inequality in the number of citations received by papers increased, and as a consequence the shortest paths in the network depend more on high impact papers later in the period. Third, the dispersion of connections between fields increased continually. Moreover, these changes did not entail a lower level of clustering of citations. Both within- and cross-field citations show a similar rate of slowly growing clustering values in all years. The latter findings suggest that domain spanning scholarly communication is partly enabled by new fields that connect disciplines.

**Significance Statement**: The constantly expanding volume of scientific research engenders specialization, which narrows the focus of research fields. Does this pattern of scientific growth prevent information from circulating between fields? Does motivation to explore new problems and combine innovations across domains counteract this process? This analysis, based on the Science Citation Index, shows that the distances in citation networks decrease from 1950 to 2018. This provides evidence that the sciences diffuse information more easily over time. The shortened distances are due to more dispersed citation activity between fields and growing centralization of citations. Despite these changes, the clustering of citations did not decrease. Co-citations are slightly more embedded over time, which suggests that cross-field ties create their own field boundaries.

# Introduction

Scientific research is conducted in specialized subfields. The division of labor helps to maintain an effective production system, which splits knowledge and expertise into manageable units. Human intelligence cannot effectively handle increasingly large volumes of information, and therefore the organization of learning and evaluation of new knowledge necessitates autonomous expert networks (1-3). As subfields grow they spawn new specialties (4). This fragmentation into subspecialties is a marked property of scientific advancement, as the number of publications has been growing exponentially since the scientific revolution (5-8).

From the individual scientist's perspective, the proliferation of new specializations is often a worrisome development. It correlates with the frustrating impression that potentially relevant literatures grow at a pace that makes monitoring relevant information impossible. More importantly, subsequent specialization confines the focus of research and education (9-11, 4). On the other hand, the motivation to innovate offsets the over-expansion of the scientific universe and stimulates cross-field communication. It is a widely held assumption that importing information from disparate fields could lead to novel ideas (12-16). Accordingly, research policy doctrines propagate interdisciplinary practices to incentivize knowledge synthesis (17), and there is evidence that interdisciplinarity is becoming somewhat more popular (18, 19). Other institutional changes may also promote the integration of distant corners of the sciences. The communication infrastructure around scientific research is improving, the importance of team science is growing (20), and universities now often establish research centers to foster interdisciplinarity and focus on applied topics (21).

Are the modern sciences becoming fragmented due to the enormous growth of scholarly output? Are the incentives to broker information balancing out this tendency, and even blurring the boundaries of specializations, as scholars suggest (21, 22)? To answer these questions, this study investigates the temporal evolution of citation networks retrieved from Web of Science (WoS), and reports the distances and clustering of these networks. The general research questions are broken down to four tractable analytical questions.

First, how did the average citation distance change in the literature? It is a common assumption in scientometrics that ideas in the sciences are disseminated through references (23) and that patterns of citations and co-citations are indicative of field boundaries and the evolution of knowledge domains (24-25). If the sciences are unable to maintain integration, the distances between the cited literatures would increase, and the diffusion of ideas would be more difficult. The opposite scenario is that the above mentioned institutional trends counterweight fragmentation, in which case the distances between publications are decreasing. Second, the trend of citation distance is perhaps influenced by changes in the distribution of citation impact. Scientific credit is allocated unevenly (26), and the connectivity of the citation network depends heavily on these high impact papers (27). Accordingly, Derek de Solla Price (6) – who first studied the exponential growth of science – speculated at the dawn of "big science" in the sixties that networks of scientists who are prominent representatives of their respective fields will integrate research findings across specializations. Third, I also track the evolution of lateral citation relations between enduring fields, based on the WoS classification scheme for subject categories. Similar to centralization, dynamics of cross-field ties provide an explanation about why distances in the literature are changing.

Finally, does the clustering of citations remain stable over time? If the connectivity between the disciplines in terms of shorter paths in the literature is indeed improving, is this accompanied by more permeable field boundaries? Network science has shown that short paths can evolve in networks without fundamentally destroying the overall clustering of the networks (28). In a similar vein, some sociologists studying interdisciplinary research suggest that boundary spanning agendas that bridge disciplines tend to form their own discipline-like fields, and that boundary spanning research does not necessarily involve new institutional forms (21, 29). To examine the clustering of scholarly communication, this study investigates the co-citation behavior over time, and quantifies the prevalence of overlaps between reference lists of papers.

**Data and Methods**

Scholarly communication is represented as bipartite (two-mode) networks, where one set of nodes is constituted of papers that are published in a sampled year, and the references of those papers constitute the second set of nodes. No edges are possible within the same set of nodes. In this study, an edge in the network is referred to as a citation. The citation connects the referencing paper in the sampled year with its references. The two sets of nodes are called the source and the target of the citation, respectively. The source node is the paper that makes references in the sampled year, and the target nodes are the references. If a source node cited another source node (i.e. it is a citation within the sampled year) the edge is still recorded as a citation by duplicating the cited source node, and representing it as a target node as well. This method ensures that no citation information is lost in the network.

Representing the citation networks in this way is justified by the widely used technique of co-citation analysis. This technique is utilized to map fields and scientific advancement (24-25). The "link" between scientific works in this perspective is a shared reference between two papers. The co-citation is a sign that the two publications share a common subject and interest. Therefore, I assume that a chain of co-citations that links two papers is a possible channel of knowledge diffusion. Taking yearly snapshots of the evolving co-citations is an indicator of how papers have been shared and utilized between researchers at a given moment.

The examined literatures are indexed in the Science Citation Index (SCI) of WoS. The SCI is a selective index, which follows the high impact journals of each field. In relation to this, the number of publications grows at a slower rate than the overall growth of scientific literature (5). Nevertheless, the SCI is a collection of important journals, which provides a good representation of scholarly communication at the research front across all the sciences. This study includes every fifth year from 1950 until 2018. The analysis is restricted to references that can be recognized as scientific periodicals, which is a common practice in bibliometrics. It is more difficult to index books, book chapters, or ephemera (e.g. editorials, comments).

The shortest path lengths in the networks were measured to appraise the distances and connectivity of research papers through their references. This quantity is also called the graph geodesic. The average shortest path between two nodes is the minimal number of steps along the edges of the network to reach one node from another node. Because the studied networks are bipartite networks, the shortest possible path between two papers in the sampled year is a co-citation (two papers co-cite a third one), which is a 2-path. To make this measure more similar to the commonly used notion of graph distance, the presented distance values are divided by two, so the minimal distance is one. These distances have been calculated between the source papers on repeated samples of 2,000 source papers selected randomly in each year. This equals 1,999,000 paths between all pairs of nodes. This sampling method was repeated thirty times for each network. The SI Appendix describes in detail the random network generating procedure used in the study.

To quantify the clustering of the network, an edge clustering coefficient ($C$) was calculated for each citation in the network. $C$ measures the density of citations between the neighborhoods (nodes connected to a focal node) of the two nodes constituting the focal citation. In short, it measures the embeddedness of citations. This is the log ratio of the number of citations between the nodes that are connected to the focal citation's source and target nodes, and the expected frequency of citations between these nodes. While the nominator is the number of citations between the source and target paper's neighborhood, the denominator is the randomly expected number of connections between the neighboring nodes given the degrees of these nodes. A strongly embedded citation passes through a high-density part of the network (i.e. the reference lists overlap) and *C* has a high value. See the SI Appendix for more information on the calculation of *C* and its relation to similar measures.

I also utilized WoS Subject Categories, which is a journal classification system that represents subdisciplines across the sciences. Subject Categories are initially assigned to journals, and subsequently to individual papers. To assign subject categories to the target papers, I used the journal list of Science Citation Index Expanded, which indexes more journals than SCI. Although this classification system is used widely to measure interdisciplinarity (3, 18, 30) and for normalizing citation impact, Leydesdorff and Bornmann (31) warn against mapping fields of science solely based on Subject Categories. For present purposes – in line with the intentions of the developers (32) – I use them as a "heuristic method" to examine enduring disciplinary boundaries.

**Results**

The number of source papers in the networks increased between 1950 and 2018 from 18 thousand to 760 thousand, while the cited literature increased more substantially from 151 thousand to 11 million (SI Appendix, Table S1). The length of the bibliographies of the publications also increased during the same time period, on average from 11 to 35.4 references. This observation has already been made by other researchers (33). As noted elsewhere (8), the citation behavior reflects the growth of published material by referencing an increasing amount of documents.

Figure 1/A demonstrates that the distances have been decreasing throughout the studied period between source papers. The decrease is less substantial until 1970, after which the rate of change is quite steady. While the average distance was 5.33 in 1950, it has been reduced to 3.18 steps, which is a 40% decrease. The mode shortest path length decreased from 5 steps to 3 steps (Figure 1/B). The chance in 1950 that two randomly selected papers are three references away from each other is 0.116, and in 2018 it increases to 0.725. These findings are robust when applying larger sample sizes to estimate the distances (SI Appendix, Figure S3). An alternative way to define the links in the network is to take into account the size of the overlap between the reference lists of source papers. In this case, the more references two papers share, the shorter the distance between the two papers. Following this weighted distance approach to appraise the changes we observe similar results (SI Appendix, Figure S4).

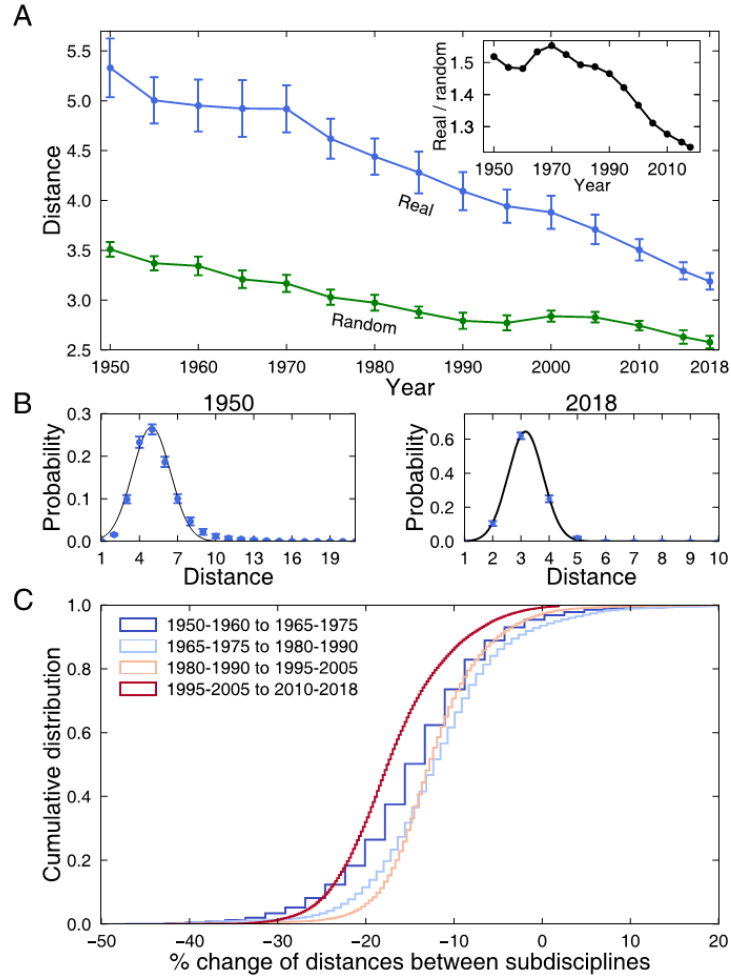

**Figure 1.** Changing distances in the citation networks and between subdisciplines (Subject Categories). (**A**) Average distance in the citation networks and in their random counterparts by time. The average distances in the real networks were calculated with repeated sampling: a random sample of 2,000 nodes was created for each network, and the sampling was repeated thirty times. For each sample the average distance was calculated between the 2,000 nodes. The figure shows the averages three SDs across the repeated samples (blue line). The green line shows the average distances and three SDs across the thirty random networks. The average distance in a single random network was calculated between 2,000 randomly selected nodes. (**B**) Distribution of distances at the beginning (1950) and at the end (2018) of the studied time period. Each marker on the figure represents the average probability of a given distance in the thirty repeated samples. Error bars are three SDs. Normal curves are fitted to the distributions. (**C**) Cumulative distribution of distance changes between subdisciplines in four time intervals. The distributions on the figure are based on the matrices in SI Appendix, Figure S5. The latter matrices on Figure S5 show the average distance between the subdisciplines. For a given three-year time period, I took the average of the corresponding cells (representing the distance between two subdisciplines). The figure here shows the cumulative distribution of the percentage change of the averages between the two time periods for each cell.

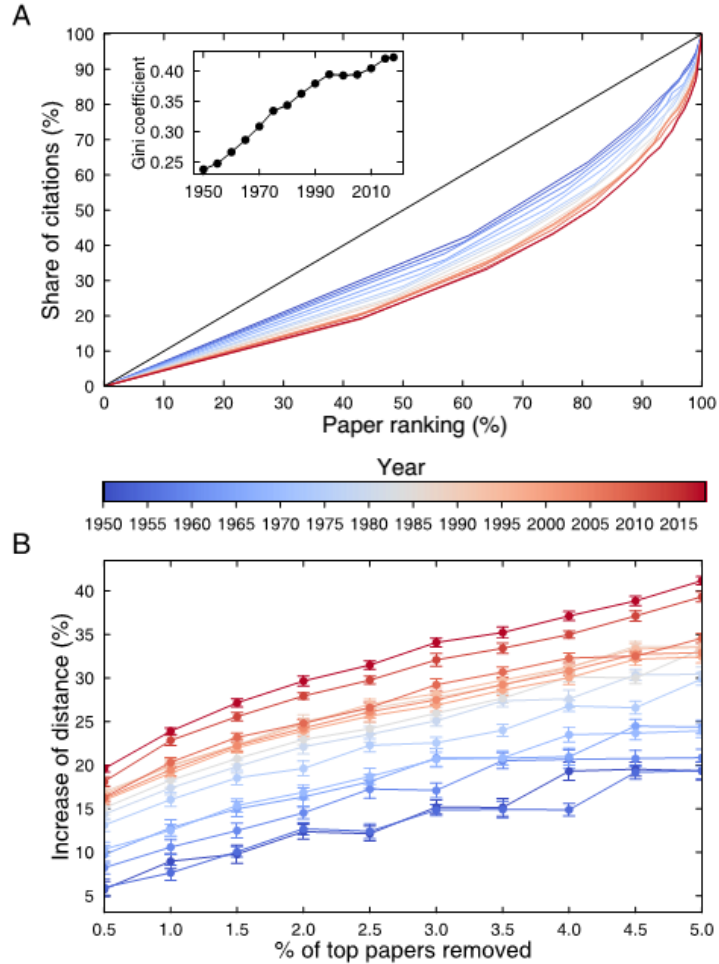

**Figure 2.** Growing citation impact inequality and its influence on the changing distances. (**A**) Lorentz curves and Gini coefficients describing the distribution of citations to target papers (i.e. citation impact). Target papers with only one citation are omitted from the distributions, because they don't affect distance, which is the subject of this study. (**B**) The increase of distances after removing the highest impact papers from the network. At each removal the average distances have been estimated on a repeated sample of 2,000 nodes. The error bars show the SDs of the average distances across the thirty repetitions.

Could it be that the longer reference lists in articles can account for the decreased distances in the literature? One way to approach this question is to compare the observed trend with the distances in random networks, which have the same degrees as the real networks. Figure 1/A shows the average distances in the random networks, which serve as baselines. The ratio of the observed distances and the baseline measures is decreasing as well, which overall suggests that the more extensive surveying of the literature later in the period would not be responsible in itself for the shortened distances.

Is the decrease of distances consistent across subdisciplines? To study the distribution of distances across subdisciplines, I sampled 10,000 source articles in each year, and measured the distances of all the source articles to this sample. From this I assembled the subdiscipline distance matrix, in which the rows and columns are the subdisciplines, and the cells indicate the average distances between the papers in the particular subdiscipline pairs. The rows are based on all source papers, while the columns represent the subdisciplines of the sampled papers. Figure 1/C shows the distribution of the percentage change of the average distance between all subdiscipline pairs across four time periods. The proportion of subdiscipline pairs in which the distances increased is between 4.8% and 0.8% in the four time intervals. SI Appendix, Figure S5 shows the distances between subdisciplines in each year, which reveals that the decrease of distances is more or less homogeneous across the subdisciplines, except that the decrease of distances between the life sciences is less dramatic. In fact, the subdisciplines within the life sciences had shorter distances throughout the whole period, which is why the decrease of distances is slightly less substantial.

The majority of the shortest paths in networks with strongly right skewed degree distributions is likely to pass through the most well-connected nodes of the network (27). Citation impact, or the degrees of the target nodes, has a distribution that closely follows the power law (34). It is therefore important to investigate closely how that distribution affects the changing distances. Figure 2/A shows the Lorentz curves of these degree distributions. The curves indicate the share of the total citations by the top x-percent of nodes. A curve runs closer to the bottom-right corner as we approach the present, which demonstrates that the inequality is growing. While in 1950 the top 4% of the papers received 13.6% of the citations, by 2018 their share grows to 24.2%.

To investigate further the effect of growing citation impact inequality on distances, a test was conducted to determine whether the robustness of the average shortest path length depends more over time on the top target papers. By removing these nodes and recording the average shortest path lengths in the resulting network one can reveal to what extent shortest paths pass through the top nodes. Figure 2/B shows that indeed this dependence increases over time. The removal of the top 0.5% target papers increases the distances in 1950 by 5.7%. In 2018, the increase after the removal is 19.56. Removing the top 5% causes a 19.3% increase of the average distance in 1950 and a 41.2% increase in 2018. Although the deletion of these nodes shifts the average distances, it does not alter the shape of distribution of distances, which generally follows a Gaussian function (SI Appendix, Figure S6).

Increased lateral connectivity between fields provides another explanation of the decreased distances. I approach this question by examing the relations between subdisciplines with subdiscipline citation matrices (SI Appendix, Figure S7), where cells show what percentage of all citations from source papers in given subdisciplines refer to target papers in any subdiscipline. The typical citation stays within a single subdiscipline (i.e. both the source and target papers are in the same subdiscipline). The average percentage in such cells decreases over time from 29.8% to 18.3% (Figure 3/A). The changes from 1950 to 1975 could be partly due to the increasing number of subdisciplines in the dataset, but by the 1990s this number was fairly stable and does not affect the results (SI Appendix, Figure S8/A).

Citations are also less concentrated into specific subdisciplines. Figure 3/B-C shows the Herfindahl-Hirschman concentration indexes (HH index) for each subdiscipline. The HH index (which is also called the Simpson index in ecology) is computed for each row of the subdiscipline citation matrix. In the current context, the index is the probability that two target papers – taken at random and cited by source papers published in a specific subdiscipline – are in the same subdiscipline. The higher this index, the more concentrated the citations are to specific subdisciplines. The figure reports two variants of the index. In the first case, the computation of the index takes into account all target subdisciplines (Figure 3/B). In the second case, it only measures the concentration among the subdisciplines that are different from the source papers's subdiscipline (Figure 3/C). Both variants of the index decrease steadily by around 60%. In other words, the predictability of the target subdiscipline of a citation based on the source paper's subdiscipline decreases over time. This is not simply due to the decrease of homophilious citations (citations with the same source and target subdisciplines), because the second variant of the index – which omits these citations – shows the same trend. These results suggest that the sciences increased their interconnectivity by developing connections and possibly new fields that bridge the subdisciplines. Further analysis shows that this trend is not affected substantially by the growth in the number of subdisciplines (SI Appendix, Figure S9). It should be noted that the distribution of all citations across the subdisciplines is becoming more even over time (SI Appendix, Figure S8/B-C). However, the decreasing concentration of citations at the level of sudisciplines is independent of this general trend (SI Appendix, Figure S8/D-E). Finally, it is also important to note that cross-subdisciplinary ties have likely been developing between subdisciplines that are already cognitively close to each other (18).

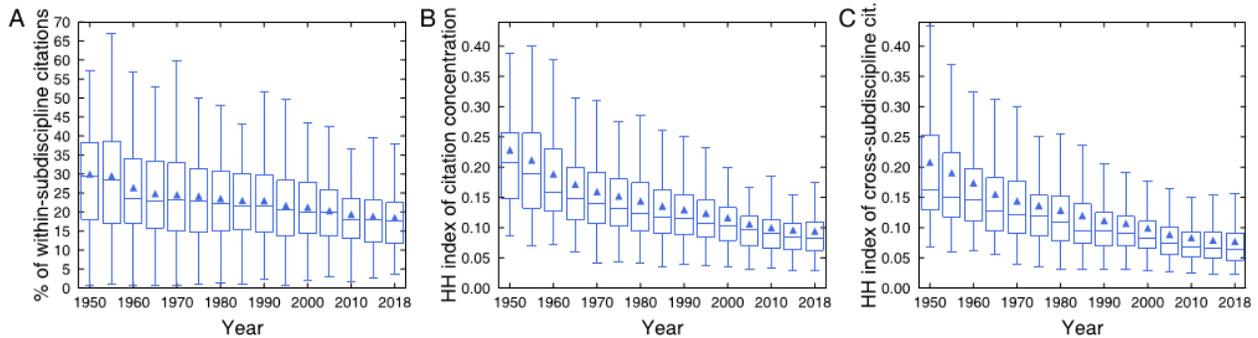

**Figure 3.** Distribution of citations between subdisciplines (Subject Categories). (**A**) Percentage of citations where both the source and target articles are in the same subdiscipline. The figure shows the distribution of within subdiscipline citations. Boxes show the interquartile range, whiskers indicate the range, the middle vertical lines are the median, and triangles are the means. A Kruskal-Wallis H-test has been performed (P < 0.001) to see if the temporal trend is statistically significant. The pair-wise comparison of the years with Welch's t-tests shows that 67 out of 105 year-pairs differ significantly from each-other at the P < 0.05 level. Therefore, the association is not driven by a few diverging years. (**B**) Herfindahl-Hirschman indexes for each subdiscipline. The HH index is a concentration measure. In the context of the present article it is the probability that two randomly chosen citations with the same source article subdiscipline have the same target discipline. The index is higher if the citations of papers originating from the specific subdiscipline disperse more to all the subdisciplines. The Kruskal-Wallis H-tests are statistically significant (P < 0.001). The Welch's t-tests indicate that the average HH indexes differ (P ≤ 0.05) in 84 out of the 105 year pairs. (**C**) While the first variant of the HH measure takes into account all citations from the given subdiscipline, the variant here only considers the distribution of the citations where the source and the target subdisciplines differ. Kruskal-Wallis H-tests are statistically significant in this case as well (P < 0.001). 83 out of the 105 Welch's t-tests are significant (P ≤ 0.05).

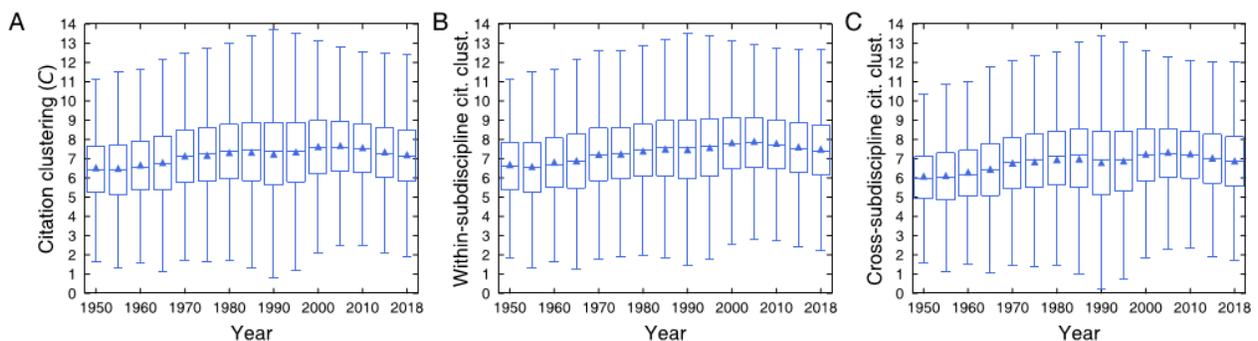

**Figure 4.** Edge clustering coefficients. (**A**) Distribution of $C$ for all citations in the networks. The pairwise comparison of the clustering coefficients with Welch's t-tests are all statistically significant (P < 0.001). (**B**) Clustering coefficients of citations situated within the same subdiscipline. (**C**) Distribution of $C$ for cross-subdisciplinary citations.

Although the data presented here provide evidence for an increasingly interconnected scholarly communication network, this trend is not accompanied by decreased clustering of the networks. The edge clustering coefficient $C$ does not follow a linear trend (Figure 4/A), and is rather stagnant. While $C$ increases between the beginning and the end of the studied period, it falls back slightly from 2005. The SI Appendix provides a more detailed examination of edge clustering. In short, the number of connections in the vicinity of the median citation increased over time, while the expected number did not change substantially, which is why $C$ shows the temporal trend described above.

Because we have seen lateral connections between fields became more prevalent, it is necessary to examine how clustering between differing subdisciplines changes. Citations with the same source and target subdisciplines have slightly higher edge clustering than those citations that connect different subdisciplines in all years (Figure 4/B-C). The trend is very similar to the overall trend of $C$, as both of these coefficients of the two sub-samples increase. In conclusion, although lateral citations were on the rise constantly, the embeddedness of citations across subdisciplines does not decrease in the studied period.

**Discussion**

Fear of fragmentation in the sciences is shared by both scientists and policy makers. The latter promote boundary spanning research that integrates branches of the sciences to respond to scientific and societal needs, which requires a unified perspective (17). Fragmentation can also lead to unnecessary parallel discoveries, resulting in non-optimal allocation of work effort (35). Enormous scholarly output, especially in the medical sciences, has recently motivated scholars of information retrieval and computational linguistics to automate information aggregation from the text of published material (36).

The evidence presented above suggests that science has become more interconnected over time, despite continuous expansion. This increased connectivity can be explained by the centralization of citations over time and the growth of cross-field communication. These conclusions are based on three main observations. First, distances decreased between almost all subdisciplines. Second, the citation impact inequality was rising and shortest paths in the citation network had an increased dependence on top papers. Third, the salience of citations between the same subdisciplines slightly decreased, and the dispersion of citations between subdisciplines increased significantly from year to year.

Finally, this increased interconnection of fields did not reduce the embeddedness of citations. While the scientific "small world" shrank further, clustering slightly increased. Cross-subdiscipline citations became more prevalent and more diverse over time. However, the average clustering of these citations remains high. It is quite conceivable that a disciplinary framework provides the organizational background for growing cross-fertilization (21, 29). These findings suggest that domain spanning scholarly communication is enabled by new fields that connect disciplines. This study provides further evidence that cross-disciplinary fields can demarcate themselves similarly to disciplines, and at the same time they can establish new bridges in the sciences.

Universities and research organizations have always been at the forefront of new communication technologies. The second half of the studied period experienced an accelerated development of digital and online indexing and abstracting services, and of electronic publishing and data sharing (37). However, the decrease of distances is steady and constant throughout this period, and no salient trend change is detectable that could be tied, for example, to the widespread use of the Internet beginning in the late 1990s. The growth of co-authorships and multi-university research collaborations shows the same even trend (20, 38).

While the findings about citation connectivity presented herein do indicate growing integration of scholarly communication, it is quite conceivable that other forms of fragmentation pose problems for knowledge synthesis. One type of fragmentation mentioned above is when the literature output on a topic is so vast that researchers cannot monitor new findings effectively. Another type of fragmentation occurs when scientists are not motivated to pursue research synthesis and instead concentrate their efforts on specialized research (4). New information infrastructures, innovative approaches for research synthesis, and research policy initiatives may overcome these difficulties in the future.

**Acknowledgements**

I thank Ronald Breiger, Erin Leahey, Loet Leydesdorff, Peter Ore, Yotam Shmargad, Gretchen Stahlman, and Mihai Surdeanu for suggestions and helpful discussions. I am also grateful to the University of Arizona's HPC services for aiding my work, and to the anonymous reviewers for helping to improve and clarify this paper. The Web of Science data are available via Clarivate Analytics.